\documentclass[aps,twocolumn,showpacs,preprintnumbers,amssymb]{revtex4}

\newcommand{\be}{\begin{eqnarray}}
\newcommand{\ee}{\end{eqnarray}}

\newcommand{\eins}{\mbox{$1 \hspace{-1.0mm}  {\bf l}$}}
\def\bea{\begin{eqnarray}}
\def\eea{\end{eqnarray}}
\def\C{\hbox{$\mit I$\kern-.7em$\mit C$}}
\def\N{\hbox{$\mit I$\kern-.3em$\mit N$}}

\def\tr{{\rm tr}}

\usepackage{dcolumn} % Align table columns on decimal point
\usepackage{bm} % bold math

\usepackage{epsfig}

\usepackage{amsmath,amsfonts,amssymb,graphics,graphicx,epsfig,color,times,bbm}

\begin{document}

%\bibliographystyle{apsrev}

%%%%%%%%%%%%%%%%%%%%%%%%%%%%%%%%%%%%%%%%%%%%%%%%%%%%%%%%%%%%%%%%%%%%%%%%%%%%%%

\title{Multipartite secure state distribution}

\author{W. D{\"u}r$^{1,2}$, J. Calsamiglia$^{1}$ and H.-J. Briegel$^{1,2}$}

\affiliation{$^1$ Institut f{\"u}r Theoretische Physik, Universit{\"a}t Innsbruck,
Technikerstra{\ss}e 25, A-6020 Innsbruck, Austria\\
$^2$ Institut f\"ur Quantenoptik und Quanteninformation der \"Osterreichischen Akademie der Wissenschaften, Innsbruck, Austria.}
\date{\today}

\begin{abstract}
We introduce the distribution of a secret multipartite entangled state in a real--world scenario as a quantum primitive. We show that in the presence of noisy quantum channels (and noisy control operations) any state chosen from the set of two--colorable graph states (CSS codewords) can be created with high fidelity while it remains unknown to all parties. This is accomplished by either blind multipartite entanglement purification, which we introduce in this paper, or by multipartite entanglement purification of enlarged states, which offers advantages over an alternative scheme based on standard channel purification and teleportation. The parties are thus provided with a secret resource of their choice for distributed secure applications.
\end{abstract}


\maketitle
%%%%%%%%%%%%%%%%%%%%%%%%%%%%%%%%%%%%%%%%%%%%%%%%%%%%%%%%%%%%%%%%%%%%%%%%%%%%%%

% Introduction: Quantum primitive in real world scenario
\section{Introduction}

In classical information theory a number of basic primitives are known
 ---among them bit commitment, coin tossing, fingerprinting, Byzantine agreement and key distribution---, which serve as building blocks for practically relevant applications. Several of these classical primitives have also been investigated in a quantum set--up. It was shown that quantum features allow one to perform certain tasks apparently impossible in a classical set--up, the most prominent example being (quantum) key distribution, which allows for unconditional secure communication. The quantum nature of states offers, however, not only additional possibilities to realize classical primitives, but also allows us to consider new primitives that are intrinsically quantum and hence do not have a counterpart in classical information theory or classical physics. An important aspect of such quantum primitives ---which has been mostly neglected in previous discussions--- is the stability of these concepts under imperfections, i.e. the realization of the task in a real--world scenario. A task that might seem trivial in an idealized scenario may become highly non--trivial or even impossible under realistic conditions, i.e. when taking inevitable noise in quantum channels and local control operations into account. We emphasize that noise is not just a practical issue, but is a fundamental limitation one has to cope with in quantum systems. 

In this paper, we will discuss a robust quantum primitive, which may be used as a basic building block for both quantum and classical security applications in situations where local and channel noise are present. Specifically, we will consider the secret creation of a spatially distributed multipartite entangled state with high fidelity. We will consider the set of all two--colorable graph states as possible target states. Two--colorable graph states include many qualitatively different types of multipartite entangled states, each of which can serve as a different resource to perform certain security tasks. In particular, they include any collection of bipartite singlet states shared between some of the parties, any type of multipartite GHZ states, so--called cluster states \cite{Ra01,Ra01b} (a universal resource for measurement based quantum computation) as well as algorithmic specific resources which allow one to implement a certain algorithm (or non--local unitary operation) by means of measurements. As has been shown recently, two colorable graph states are equivalent to codewords of Calderbank--Shor--Steane (CSS) codes \cite{Ch04}. It may be of interest to the end--users that the state remains secret to any third party, i.e. nobody else knows which specific  resource  they share (and hence which tasks they are capable to perform). The information about the state can even remain unknown to the parties themselves, being only disclosed to a single end-user ( e.g. a trusted party or dealer). 
This is relevant in the implementation of distributed secure quantum applications over noisy communication channels, which count with an additional `adversary', the eavesdropper  (that should be considered as the noise source in the channel), and that might collaborate with the untrustful parties. There are standard purification protocols \cite{Be96,De96,Mu99,Sm00,Du03,As04,Ch04} used to reduce noise levels and factor out any possible eavesdropper even in the presence of noise \cite{As01}. However, these protocols can not straightforwardly be accommodated to account for possible untrustful parties ---which actively participate in the purification protocol---, and for their possible complicity with eavesdropper.

In this paper, we present ways to achieve the secure distribution of the aforementioned graph states under realistic conditions, i.e. noisy quantum channels and imperfect local control operations. After fixing notation and describing the scenario for secret state distribution in Sec. \ref{Def}, we present three possible solutions to the problem in Sec. \ref{3sol}. The first approach, described in Sec. \ref{Channelpury}, is based on channel purification and teleportation, while the second and third approach deal with direct multipartite entanglement purification. The second approach, described in Sec. \ref{Direct}, uses blind multipartite entanglement purification, which we introduce in this paper. In the third approach (see Sec. \ref{Enlarged}), the security is guaranteed by purifying an enlarged entangled state. We summarize and conclude in Sec. \ref{Summary}.

\bigskip

% Introduction to Graph states and CSS codes, possible applications
\section{Definitions and description of the scenario}\label{Def}

\subsection{Two--colorable graph states}
\begin{figure}[ht]
\begin{picture}(230,130)
\put(-5,0){\epsfxsize=230pt\epsffile[50 606 388 792]{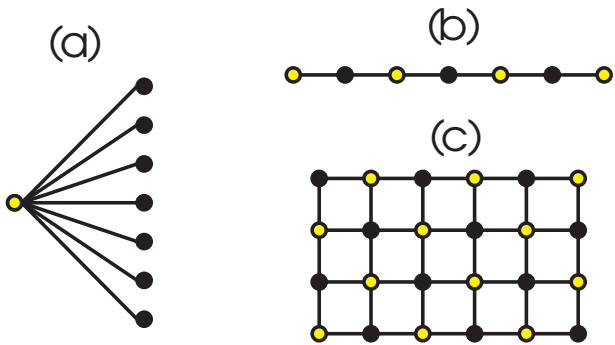}}
\end{picture}
\caption[]{\label{twoC} Examples of two--colorable graphs which correspond to (a) GHZ state; (b) linear cluster state; (c) two--dimensional cluster state. Vertices with same color are not connected by edges.}
\end{figure}

We start by defining two--colorable graph states. A graph $G$, given by a set of $N$ vertices $\{1,2,\ldots ,N\}$ connected in a specific way by edges $E$, is called two--colorable if there exists two groups of vertices, $A$,$B$ such that there are no edges inside either of the groups, i.e. $\{k,l\} \not \in E$ if $k,l \in A$ or $k,l \in B$ (see Fig. \ref{twoC}). 
To every such graph there corresponds a basis of $N$--qubit states $\{|{\bm \mu}\rangle_G\}$, where each of the basis states $|{\bm \mu}\rangle_G$ is the common eigenstate of $N$ commuting correlation operators $K_j^G$ with eigenvalues  $(-1)^{\mu_j}$, ${\bm \mu} = \mu_1\mu_2\ldots \mu_N$. That is, they fulfill the set of eigenvalue equations 
$K_j^G |{\bm \mu}\rangle_G = (-1)^{\mu_j}|{\bm \mu}\rangle_G$, $j=1,\ldots,N$. The correlation operators are uniquely determined by the graph $G$ and are given by 
\be
K_j^G= \sigma_\alpha^{(j)} \prod_{\{k,j\} \in E} \sigma_\alpha^{(k)},
\ee
where $\alpha=x$ [$\alpha=z$] if $j \in A$ [$j \in B$] respectively and $\sigma_\alpha^{(k)}$ denotes the application of the corresponding Pauli operator by party $k$. Note that the so-defined graph states are identical to the usual graph states, as introduced in \cite{Ra01b} and e.g. used in Ref. \cite{Du03,As04}, up to local Hadamard operations performed on all particles in $B$. As has been shown recently \cite{Ch04} they are in fact equivalent to codewords of the CSS codes. We also remark that the correlation operators $\{K_j\}$ are the generators of a group which is often called stabilizer of the state $|\bm 0\rangle_G$, and the corresponding description in terms of the stabilizers is also referred to as the stabilizer formalism. 

We will also consider mixed states $\rho$, which for a given graph $G$ can be written in the corresponding graph state basis $\{|{\bm \mu\rangle_G}\}$, $\rho =\sum_{{\bm \mu},{\bm \nu}} \lambda_{{\bm \mu}{\bm \nu}} |{\bm \mu}\rangle\langle {\bm \nu}|$. We will often be interested in fidelity of the mixed state, i.e. the overlap with some desired pure state, say $|{\bm 0}\rangle_G$, $F= \langle {\bm 0}| \rho |{\bm 0}\rangle$.

%Description of the scenario:
\subsection{Description of scenario}
We consider a central party $C$ (the company), which is connected via noisy quantum channels to $N$ spatially separated local agents $A_j$, $j\in\{1,2,\ldots,N\}$. The company offers the service to spatially separated customers to deliver upon request any multipartite entangled state chosen from the set of two--colorable graph states, specified by the graph $G$ and the (basis) index $\bm \mu$, with chosen fidelity $F=1-\epsilon$. The state is delivered by the local agents $A_j$ to the end--users $E_j$. The company guarantees as a special security service that
% in case of successful delivery
the local agents, a potential eavesdropper, or any other party different than the one who placed the order cannot learn any information about the chosen state. 
%The security requirement can further be strengthened in the sense that also in the case of non--%successful delivery the identity of the state should remain unknown. We will later present a modified
% protocol that enjoys this stronger security conditions. 

\section{Secret state distribution: Possible solutions}\label{3sol}

We remark that in the case of noiseless quantum channels between $C$ and the local agents, the described task can be trivially achieved by creating locally at $C$ a single copy of the requested state and distributing it to the local agents. The single copy does not allow the local agents and a potential eavesdropper to learn information about the graph $G$, even if they decide to cooperate and perform measurements. This follows from the fact that the basis index ${\bm \mu}$ is unknown and random from the point of view of the local agents, which implies that for any chosen graph $G$ the ensemble of states $\{|{\bm \mu}\rangle_G\}$ forms the identity and ensembles corresponding to different graphs $G$ are hence indistinguishable.
However, if (as in a realistic scenario) the quantum channels connecting the central station $C$ with the local agents $A_j$ are noisy, the task becomes non--trivial. In this case, the state obtained by the local agents $\rho_{\bm A}$ will be mixed and the fidelity will be below the desired one.

%SOLUTION 0: Entanglement purification - bipartite and multipartite

\begin{figure}[ht]
\begin{picture}(230,110)
\put(10,0){\epsfxsize=230pt\epsffile[23 368 582 603]{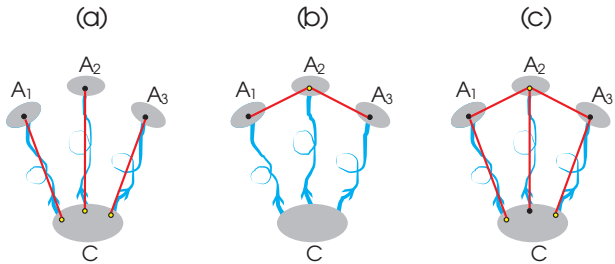}}
\end{picture}
\caption[]{\label{Setup} Set--up for secure distribution of multipartite entangled states based on (a) Channel purification {\bf (i)}; (b) direct purification of multipartite entangled states {\bf (ii)}; (c) purification of enlarged entangled states {\bf (iii)}.}
\end{figure}

Realistic applications must be designed to cope with two sources of errors, namely noisy channels and noisy local operations.
It is well known that problems arising from the noisy channels can be overcome
using  {\em entanglement purification}. In a multi-partite scenario we will thus opt for one of the following approaches depending on the particular conditions (see Fig. \ref{Setup}):
\begin{itemize}
\item[{\bf (i)}] Channel purification (see Sec. \ref{Channelpury}): The channels itself may be purified. This can be accomplished by sending parts of maximally entangled singlet states through each individual channel from $C$ to $A_j$ and creating from the resulting multiple copies of the noisy bipartite entangled states a few copies with high fidelity by a sequence of local operations and classical communication (LOCC), e.g. using one of the entanglement purification protocols of Refs. \cite{Be96,De96}. This high--fidelity entangled states can then be used for teleportation \cite{Be93} and serve as a purified channel, thereby allowing for the distribution of arbitrary multipartite entangled states.

\item[{\bf (ii)}] Direct multiparty entanglement purification (see Sec. \ref{Direct}): The resulting multipartite mixed states $\rho_{\bm A}$ can be purified. That is, several copies of the multipartite mixed states $\rho_{\bm A}$ are produced by distributing the locally created graph state through noisy channels to the local agents $A_j$. A suitable sequence of LOCC, e.g. the purification protocol for two--colorable graph states introduced in Ref. \cite{Du03}, allows us to create few copies with (arbitrary) high fidelity. 

\item[{\bf (iii)}] Purification of enlarged states (see Sec. \ref{Enlarged}): One can purify an enlarged entangled state, i.e. a (graph) state that is entangled with additional particles that are kept at the central station, while the remaining particles are sent through the noisy channels. The resulting mixed state $\rho_{{\bm A}{\bm C}}$ is purified by means of direct multipartite entanglement purification as in {\bf (ii)}, and the desired state shared among the parties $A_j$ can be created from the purified state (ideally given by $|\Psi_{{\bm A}{\bm C}}\rangle$) by means of local measurements in $C$.
\end{itemize}

On top of the noise from channels, there is also noise in the local control operations. In this case, it turns out that entanglement purification is still possible, although the minimal required fidelity $F_{\rm min}$ (i.e. the maximal acceptable channel noise) as well as the maximal reachable fidelity $F_{\rm max}$ of the purified states is limited by the amount of noise in the local control operations \cite{Br98,Du03,As04} and in fact strongly depends on the entanglement purification protocol that is used. It was shown in Ref. \cite{As04} that  direct multipartite entanglement purification offers advantages over protocols based on bipartite purification. In particular, for graph states of small degree and a generic noise model for local operations, the reachable fidelity for multipartite states created by bipartite entanglement purification (i.e. channel purification) and teleportation turns out to be {\em smaller} than the fidelity of the states created by direct multipartite entanglement purification, i.e. $F_{\rm max}^{(ii)} > F_{\rm max}^{(iii)} > F_{\rm max}^{(i)}$. This is true for all values of the local noise (see Fig. 7 in \cite{As04}). The upper fixed points $F_{\rm max}$ only depend on noise in local control operations and on the entanglement purification protocol. 
On the other hand, the minimal required fidelity $F_{\rm min}$ (which again depends on the purification protocol and noise in local control operations) puts limits on the maximal acceptable channel noise. The minimal required fidelity $F_{\rm min}$ fulfills for uncorrelated channels $F_{\rm min}^{(i)} < F_{\rm min}^{(iii)} < F_{\rm min}^{(ii)}$, while in the case of correlated channels the situation can also be the other way around. In general, each of the three schemes {\bf (i-iii)} has its own advantages. There are parameter regimes (noise level of local operations, channel noise) for which a certain scheme allows one to create an entangled state with sufficiently high fidelity, while the other two schemes fail. In particular, it can happen that a multipartite state cannot be produced with required fidelity $F=1-\epsilon$ as requested by customers when using method {\bf (i)}, while a protocol based on direct multipartite entanglement purification {\bf (ii)} or {\bf (iii)} enables one to reach the required fidelity. These facts are our main motivation to provide alternative methods to {\bf (i)} to accomplish the secret creation of a multipartite entangled state, that are based on direct multiparty entanglement purification. %\cite{notepurif}. 

%SOLUTION 1: Purification of noisy channels + teleportation - disadvantages

%\bigskip
\subsection {Channel purification}\label{Channelpury}

For perfect local control operations, approach {\bf (i)} immediately leads to a protocol to create arbitrary multipartite entangled states with high fidelity in such a way that local agents and possible eavesdroppers do not learn any information about the created state. The purification of the channel does not contain any information about the state, and the presence of eavesdroppers can be excluded by checking the resulting singlet states (e.g. by testing violation of Bell's inequality or, simpler, by measuring the expectation values of the correlation operators $\langle K_j \rangle$, $j=1,\ldots ,N$) \cite{notecorrelationoperators}. After successful teleportation the local agents only possess a single copy of an unknown state in a random basis. Again, as in the case of noiseless channels, the ensembles corresponding to different graphs $G$ are indistinguishable. The teleportation process may even be postponed until the local agents deliver (several copies of) the purified singlets to the end--users (who may then check themselves the validity of the singlets together with $C$ by testing a random subset) before teleporting the requested multipartite state. This circumvents any possible attempts of the local agents or eavesdroppers to learn information about the state or to corrupt the delivered state at a later stage. 
Security requirements are fulfilled even in the presence of imperfect control operations, as no information regarding the finally distributed state can be learned.

% Purification based on multipartite purification

\subsection{Direct multipartite entanglement purification}\label{Direct}

We now turn to the second scenario {\bf (ii)}, the direct purification of noisy multipartite entangled states, namely two--colorable graph states. For simplicity of the analysis, we consider in the following noiseless local operations. This restriction is, however, not crucial and will in fact be dropped later on. We start by summarizing the main steps involved in any multipartite purification protocol \cite{Du03,As04}:
(i) Depolarization of the mixed state $\rho$ to standard form $\rho_G$ diagonal in the basis of entangled states to be purified; (ii) Local operations on two (or more) copies of the state $\rho_G$ in such a way that information about the first state(s) is transferred to the last one; (iii) Measurement of the last state to retrieve this information, public announcement of the measurement outcomes, and keeping or discarding the remaining states depending on the measurement outcomes. 

It is easy to see that information about the state $|{\bm \mu}\rangle_G$ to be purified, both about the graph $G$ and the used basis ${\bm \mu}$, can be learned in various ways by all local agents involved in such a protocol. In the depolarization process (i), elements of the stabilizer of $|{\bm \mu}\rangle_G$ (all possible products of correlation operators $K_j^G$) are applied \cite{Ch04,Du03,As04}, which reveal information about the graph $G$. The local operations and measurements performed in the second step provide knowledge about the structure of the graph. For example, in the recurrence protocol of Ref. \cite{Du03} the two--coloring of the graph is revealed. The information about the measurement outcomes together with the fact that a particular outcome is interpreted as a successful step also allows the parties to obtain information about the graph. The statistics of measurement outcomes in several steps of the protocol can be used to learn both $G$ and ${\bm \mu}$, as for all possible graphs the expectation values of $K_j^G$ can be calculated and the corresponding histograms can be analyzed.  
%For example, at later stages of the protocol (i.e. fidelity already close to 1), one finds for the graph in question expectation values $<K_j^G> \approx \pm 1$ for all $j$ (where the sign provides information about the basis ${\bm \mu}$, while for all other graphs $G'$ some of the expectation values will be clearly $|<K_j^{G'}>| < 1$.
 %Note that even from the success probabilities the agents can learn information about the graph. 
Finally, since in principle a large number of copies of the state are available, (some of) the local agents can decide to perform state tomography on a sub--ensemble, thereby learning all information about the state they should purify. That is, when using any (known) direct multiparticle entanglement purification protocol, the secrecy of the state to be purified is not guaranteed.

%SOLUTION 2: blind entanglement purification
%\medskip

\subsubsection{Blind purification} 

In the following we will present a modified multipartite entanglement purification protocol where it is impossible for all involved local agents to learn any information about the state being purified, even if {\em all} of them decide to collaborate and even if they conspire with a possible eavesdropper who is in full control of all quantum channels between $C$ and $A_j$. 
%The security is not compromised even if the agents try to learn information at the expense of being caught. 
The central station $C$ coordinates the action of the local agents and is the only party (besides the customer that placed the order) that knows the graph state which is being purified. The customers' order consists in the graph $G$ and the desired basis index ${\bm \mu}$, as well as an additional secret random bit string of suitable size \cite{notekey}. After receiving the order, the central party prepares $M$ two--colorable graph states corresponding to the graph $G$ and chooses the basis indices $|{\bm m_i}\rangle_G$ randomly from  a uniform distribution. For any graph $G$ the set of graph states is complete, i.e. $\sum_{\bm m} |{\bm m}\rangle_G\langle {\bm m}| =\eins$.  This implies that the completely mixed ensemble of states corresponding to two different graphs $G_1$ and $G_2$,  $\{|{\bm m_i}\rangle_{G_1}\}_{i=1,2\ldots ,M}$ and $\{|{\bm m_i}\rangle_{G_2}\}_{i=1,2\ldots ,M}$, are indistinguishable, as the corresponding density operator (i.e. the proper description of the ensemble for any observer not possessing the information about ${\bm m_i}$) is the completely mixed state ($\propto\!\!\eins$) \cite{noteensemble}. The states $|{\bm m_i}\rangle$ are then distributed through noisy channels to the local agents $A_j$. Purification of these states takes place by a protocol (see below) where no information about the indices ${\bm m_i}$ or the graph $G$ is revealed, which ensures that at any stage of the protocol the ensemble of states corresponding to different graphs remain indistinguishable by the local agents. Finally the state $|{\bm \mu}\rangle_G$ is delivered to the end--users.

% Depolarization of channel instead of Depolarization of state - for all Graphs

The first step in the purification protocol is to ensure that the mixed state $\rho$ ---
which arises after sending 
the state $|{\bm m}\rangle_G$ through the noisy channels ${\cal E}_1,{\cal E}_2,\ldots ,{\cal E}_N$---  is diagonal in the specific graph state basis $\{|{\bm n}\rangle_G\}$. In standard purification protocols this is enforced by depolarizing the transmitted state. Instead, here we depolarize  the channels ${\cal E}_j$ to a Pauli--diagonal form \cite{Du04}.  
This can be accomplished by applying probabilistically the operators $\sigma_i$ [$\sigma_i^{T}$] before [after] sending a particle through the channel in a correlated way (here $\{\sigma_i\}_{i=0,\ldots,3}=\{\eins,\sigma_x,\sigma_y,\sigma_z\}$).  Given a general initial channel  ${\cal  E}\rho= \sum_{i,j=0}^3 \lambda_{ij} \sigma_i \rho \sigma_j$ the resulting depolarized channel ${\cal \tilde E}$ is then given by ${\cal \tilde E}\rho= \sum_{j=0}^3 \lambda_{jj} \sigma_j \rho \sigma_j$.  Moreover, the action of a Pauli operator in party $A_k$ on a graph state results in another graph state: $\sigma_i^{(A_k)}|{\bm m}\rangle_G\langle {\bm m}|\sigma_i^{(A_k)}=  |{\bm m} \oplus {\bm n_i^G}\rangle_G\langle {\bm m\oplus {\bm n_i^G}}|$, where ${\bm n_i^G}$ depends on the neighborhood of particle $A_k$ specified by the graph $G$ \cite{As04}. 
That is, sending $|{\bm m}\rangle_G$ through the depolarized channels also results ---independently of the graph $G$--- in the creation of a graph--diagonal state in this specific graph state basis \cite{notedepol}. Additionally, by writing  $|{\bm m}\rangle_G= 
\sigma_z^{{\bm m}_A} \sigma_x^{{\bm m}_B}|{\bm 0}\rangle_G$ ---where ${\bm m}=({\bm m}_A,{\bm m}_B)$ and $\sigma_i^{{\bm m}_A} $ denotes the action of $\sigma_i$ on all parties $j \in A$ for which ${m_j}=1$, we find that if
\be
\rho_{\bm 0} = {\cal E}_1\ldots {\cal E}_N |{\bm 0}\rangle_G\langle{\bm 0}| = \sum_{\bm n} \lambda_{\bm n} |{\bm n}\rangle_G\langle {\bm n}|,\label{rho0}
\ee
is the state resulting from sending $|{\bm 0}\rangle_G$ through the channels, then the resulting state $\rho_{\bm m}$ when sending a different basis state $|{\bm m}\rangle_G$, is given by a simple basis shift of $\rho_{\bm 0}$, i.e. 
\be
\rho_{\bm m} = {\cal E}_1\ldots {\cal E}_N |{\bm m}\rangle_G\langle{\bm m}| = \sum_{\bm n} \lambda_{\bm n} |{\bm n}\oplus {\bm m}\rangle_G\langle {\bm n}\oplus{\bm m}|.\label{rhom}
\ee
These are non--trivial properties of states diagonal in a graph state basis that follow from the description of these states in terms of their stabilizers and from the  commutation relations of the Pauli matrices. At this stage the local agents $A_j$ thus share the $M$ states $\{\rho_{\bm m_i}\}_{i=1,\ldots,M}$ \cite{notechannel}. 

We now introduce modified multipartite recurrence protocols $P1'$, $P2'$ that give the same performance as the protocols $P1,P2$ of Ref. \cite{Du03,As04}, but operate on states $\rho_{\bm m},\rho_{\bm n}$ (which are obtained by a basis shift from $\rho_{\bm 0}$ according to Eq. \ref{rhom}) and fulfill our security requirements. We start with protocol $P1'$, where in a first step local CNOT operations \cite{noteCNOT} with the particles of the first [second] state acting as source [target] respectively are applied by all parties. It is easy to check that the action of such multilateral CNOT operations is given by
\be
|{\bm m_A},{\bm m_B}\rangle |{\bm n_A},{\bm n_B}\rangle \rightarrow |{\bm m_A},{\bm m_B}\oplus {\bm n_B}\rangle |{\bm m_A}\oplus {\bm n_A},{\bm n_B}\rangle.
\ee
All particles of the second state are then measured in the eigenbasis $\{|0\rangle_x,|1\rangle_x\}$ of $\sigma_x$, yielding results $(-1)^{\xi_i}$ which are publicly announced. From $\xi_i$ the expectation values of all correlation operators $K_j$ with $j\in A$ can be calculated at the central station $C$. The result is taken as a successful purification step if the calculated expectation values correspond to ${\bm m_A}\oplus {\bm n_A}$. In this case, one finds that the resulting state is again diagonal in the graph state basis, with new coefficients
\be
&&\tilde\lambda_{{\bm \gamma_A}\oplus{\bm m_A},{\bm \gamma_B}\oplus{\bm m_B}\oplus{\bm n_B}} =\nonumber\\
&&\frac{1}{2K}
\sum_{\{({\bm \nu_B}, {\bm \mu_B}) | {\bm \nu_B} \oplus{\bm \mu_B}={\bm \gamma_B}\}} \lambda_{{\bm \gamma_A},{\bm \nu_B}}\lambda_{{\bm \gamma_A},{\bm \mu_B}},\label{mapP1}
\ee
where $K$ is a normalization constant. Note that these coefficients are exactly the same as in the situation where we apply the original protocol $P1$ to two copies of $\rho_{\bm 0}$, only the basis of the resulting state is shifted by $({\bm m_A},{\bm m_B}\oplus{\bm n_B})$ (as done in Eq. \ref{rhom}). 
%That is, the resulting state is given by $\rho_{{\bm m_A},{\bm m_B}\oplus{\bm n_B}}$ with new coefficients $\tilde \lambda$.
 Similarly, the protocol $P2'$ consists of local CNOTs in the opposite direction, followed by measurement of all particles of the second state in the eigenbasis $\{|0\rangle_z,|1\rangle_z\}$ of $\sigma_z$. From the measurement outcomes the expectation values of all correlation operators $K_j$, $j\in B$ can be calculated at the central station $C$. The result is interpreted as a successful step if the calculated expectation values correspond to ${\bm m_B}\oplus {\bm n_B}$. Again, one can determine the action of this protocol on two states $\rho_{\bm m}\otimes \rho_{\bm n}$ and finds that the resulting state is up to a basis shift $({\bm m_A}\oplus{\bm n_A},{\bm m_B})$ the same as the one obtained by the protocol $P2$ applied to $\rho_{\bm 0}^{\otimes 2}$. The purification procedure takes place by an alternating application of protocols $P1',P2'$, where at each step the central station randomly chooses the pairs to combine. In particular, also pairs resulting from unsuccessful purification rounds are further processed. We remark that this substantially reduces the yield of the entanglement purification protocol as compared to the original scheme. The reachable fidelity and error thresholds, however, are exactly the same as of protocols $P1,P2$ investigated in Ref. \cite{Du03,As04}. The only difference to the original scheme for the central party $C$ is that it has to keep track of the basis shift for each state and modify the decision about successful/unsuccessful purification steps (which are not publicly announced) accordingly. We remark that in order to determine the basis shifts for final copies, knowledge of the initial basis shifts $\{{\bm m_i}\}$ and the history of the purification procedure is required for each state, i.e. ${\bm n_i}=f_i({\bm m_i})$. At the end of the procedure, additional independent random basis shifts are applied to all states (this can be accomplished by letting the local agents apply appropriate Pauli operators), where for a specific copy, say $1$, resulting from a successful purification branch (i.e. all purification steps successful) a basis shift ${\bm n_1} \oplus {\bm \mu}$ is performed. This guarantees that copy 1 is in the basis ${\bm \mu}$ and hence corresponds to the required state. The order of the copies is randomly chosen by $C$ in such a way that copy 1 is located at position specified by the additional secret bit string which is shared with the end--users. All states (also the ones resulting from unsuccessful purification steps) are then handed over to the end--users. This ensures that it remains unknown to all involved parties (except $C$ and the end--users) which of the copies correspond to successful branches. Finally, all states but copy 1 are measured, either in the eigenbasis of $\sigma_x$ or $\sigma_z$ and the measurement results are publicly announced. This allows the central station to check the trustfulness of the local agents, as e.g. for states corresponding to a successful purification procedure the expectation values of the correlation operators $K_j^G$ and hence the fidelity of the produced states can be determined by $C$ \cite{notecorrelationoperators}. Depending on the results of this verification procedure, $C$ uses another shared random bit to secretly announce whether the remaining copy can be used for the desired security application. We note that distrustful local agents or eavesdropper can always prevent a successful generation of the desired state, e.g. by simply not taking place in the purification procedure as requested or by adding additional noise. However, if the state passed the verification procedure it can be guaranteed that it has the required fidelity and that the local agents have not learned information about the target state. 

\subsubsection{Security of direct multiparty entanglement purification}
We now discuss the security of this modified protocol by analyzing the scheme from the point of view of the local agents (or eavesdropper). We first remark that the entanglement purification protocol is carefully constructed in such a way that absolutely no information about the state to be purified is revealed during the protocol. In contrast to original protocols $P1,P2$, the protocols are symmetric, i.e. no information about the two--coloring of the graph is revealed. The randomly chosen basis states $\{{\bm m_i}\}$ guarantees that at any stage of the protocol no information about the structure of the graph can be learned from measurement statistics, as any statistics is randomized due to random basis shifts. Even the information of whether a given purification step is successful or not is kept secret. This is guaranteed on the one hand by randomly combining copies of the state after each purification round and on the other hand by delivering the whole ensemble of states rather than a single copy (which the local agents would otherwise identify as the resulting state of several successful purification rounds). Although it seems unlikely that such a tiny amount of information could be used by the local agents and eavesdropper to learn about the identity of the graph state, it can not be excluded that there exists a strategy which allows them to make use of this information. One may e.g. imagine that by means of a graph state analyzer they were able to learn information about the basis index ${\bm m_i}$ for this specific state, which ---together with the information that the state results from several successful purification rounds--- could then be used to exclude the graphs that are incompatible with the measurement outcomes obtained in  the purification protocol. Hence, we have modified the purification protocol in the way described above, which guarantees that {\em no} additional information can be learned from the protocol itself. In fact, from the point of view of the local agents the protocol they should perform (as well as all measurement outcomes etc.) is the same for all graph states. It follows that the secrecy of the graph state to be purified is still guaranteed by the fact that the corresponding ensembles of states for two different graph states are indistinguishable, as they are both described by the identity. Naturally, this still holds if one considers the transmitted ensemble of states  $\{\rho_ {\bm m_i}\}$ instead of $\{{\bm m_i}\}$: it is safe to attribute the noise in the channel to the actions of an eavesdropper and/or the agents. 

We finally remark that unconditional security can only be guaranteed if we keep the information whether the required state has been successfully created or not secret. Otherwise there exists an (indirect) strategy for a possible eavesdropper to learn about the graph: by varying the noise in the channel (we assume that Eve has complete control over the channel), Eve together with the local agents can prevent the purification of certain states (e.g. states with high degree) as for these states the channel noise (or noise in local control operations) is above the threshold value where purification is possible. As the threshold values depend on the graph, knowing whether the purification procedure was successful or not, would provide Eve with some information about the graph,  e.g. about its (local) degree. This is just a single bit of information which might well be negligible in many cases as there exist exponentially many different graph states which are potential target states. By keeping the information of whether the procedure was successful or not secret, we prevent Eve from learning even this single bit of information. The validation stage in our protocol only serves to detect possible attempts of the local agents to prevent the production of the state with required fidelity. The secrecy of the states is guaranteed by other means (the random basis shifts). In principle, one may use the validation stage also to detect possible attempts of Eve or the local agents to gain information about the final graph, which clearly leads to corruption of the states and hence to reduced fidelity. In approach {\bf (iii)} we will discuss a strategy which guarantees security even when the success or failure of the protocol is publicly known. The strategy of {\bf (iii)} can immediately be adopted to the protocol described above.

The influence of noisy local operations can easily be analyzed. From the point of view of the central station $C$, we essentially have entanglement purification protocols $P1,P2$ with imperfect means with the corresponding properties discussed in Refs. \cite{Du03,As04}. In particular, entanglement purification is still possible even for errors of local operations at the order of (several) percent. Only the the reachable fidelity of the target state is limited. From the point of view of the local agents, the additional noise in their operations will not enable them to learn more information about the state to be purified than they would be able to learn if their operations are noiseless (they still would have to distinguish between two ensembles of states which are both ---from their point of view--- the identity at all stages of the protocol). 
Thus the produced state remains secret and the required procedure can be followed in a real--world scenario. 
% Perhaps  in summary?
The secrecy of the states at all stages is guaranteed by the random basis shifts $\{\bm m_i\}$, which on the one hand ensure that the input ensemble is the maximally mixed state, and on the other hand that measurement statistics and any association of specific outcomes with the graph--structure during the protocol itself are randomized. The second statement follows from the non--trivial property of two--colorable graph states that basis shifts can be effectively propagated through both the channels and the purification protocols, resulting in output states with shifted bases.

%SOLUTION 2: Purification of enlarged state

\subsection{Purification of enlarged states}\label{Enlarged}

We now turn to scenarion {\bf (iii)}, the purification of enlarged states. Instead of creating the desired graph state $|{\bm \mu}\rangle_G$ directly and sending several copies through the noisy channels to the local agents, the central station can also create enlarged graph states $|{\bm 0}\rangle_{\tilde G}$ of $2N$ qubits in such a way that each vertex of the initial graph is connected to an additional, independent vertex. That is, if $G=(V,E)$ is the graph of vertices $\{1,2,\ldots ,N\}$, then the graph $\tilde G$ corresponding to the enlarged state is given by vertices $\{1,2,\ldots ,2N\}$ with edges $\tilde E = E \cup \{(k,k+N)\}$ with $k=1,2,\ldots ,N$. The qubits corresponding to vertices 1 to $N$ are sent through the noisy channels to the local agents, while qubits $N+1$ to $2N$ are kept by the central party $C$. The bases randomization of initial states done in approach {\bf (ii)} is here replaced by additional quantum correlations between $C$ and $A_j$. In principle, $C$ could introduce random basis shifts on the states of $A_j$ by performing suitable measurements (reducing the protocol to the one discussed in {\bf (ii)}). However, the quantum correlations are more powerful than classical correlations, which allows us to further simplify the purification protocol. The purification takes place by a multipartite protocol and hence offers advantages (e.g. higher reachable fidelity) as compared to scheme {\bf (i)} based on channel purification. When compared to scheme {\bf (ii)}, one finds a higher robustness of the states against channel noise for uncorrelated channels. 
It is easy to check that the state $|{\bm 0}\rangle_{\tilde G}$ can be written as 
\be
\label{psi0}
|{\bm 0}\rangle_{\tilde G} = \frac{1}{\sqrt{2^N}}\sum_{\bm m} |{\bm m} \rangle_G \otimes |{\bm m}\rangle,
\ee
where $|{\bm m} \rangle_G$ is a graph state corresponding to graph $G$ of particles 1 to $N$, while $|{\bm m}\rangle = |m_1m_2 \ldots m_N\rangle$ with $m_j \in \{0,1\}$ are orthogonal product states of particles $N+1$ to $2N$ and the sum runs over all possible binary vectors ${\bm m}$. We have that $|m_k\rangle$ denotes the eigenstate with eigenvalue $(-1)^{m_k}$ of the operator $\sigma_z$ if $k \in A$ [$\sigma_x$ if $k \in B$], where $A$,$B$ correspond to the two--coloring of the graph $\tilde G$. 

The resulting noisy graph states are then purified using multipartite entanglement purification protocols $P1', P2'$ described above, which can be applied because the graph $\tilde G$ is two--colorable whenever the initial graph $G$ is two--colorable. While the local agents publicly announce their measurement outcomes, the central party $C$ keeps its measurement outcomes secret. The expectation value of each correlation operator $K_j$ ---which determine whether a purification step is successful or not--- depends on the measurement outcome in particle $j$ and its neighbors. For $j>N$, particle $j$ itself is held by $C$, while for $j<N$ the neighboring particle $(j + N)$ is held by $C$, which implies that the expectation value of $K_j$ $\forall j$ can not be determined by the local agents. The additional qubits held at $C$ act as randomizer for all measurement outcomes of the local agents. After several successful purification rounds (where in this case always two pairs resulting from a successful purification round can be combined), all but a single copy is measured in eigenbasis of $\sigma_x$ or $\sigma_z$, where again only the measurement outcomes of local agents are announced. From the measurement outcomes, $C$ can calculate expectation values for the correlation operators $K_j$ and hence verify whether the required fidelity of the states was achieved. If the states passed the validation step, i.e. the required fidelity is achieved, the local agents hand the particles of the remaining copy over to the end--users. Finally, all qubits in $C$ of the remaining copy are measured in the eigenbasis of $\sigma_z$ [$\sigma_x$] if they are in $A$ [$B$], where $A,B$ corresponds to the two--coloring of the graph $\tilde G$. If the final states were pure, i.e. $|{\bm 0}\rangle_{\tilde G}$, it follows from Eq. \ref{psi0} that for measurement outcome corresponding to ${\bm m}=m_1m_2 \ldots m_N$ with $m_j \in \{0,1\}$, the resulting state shared by the end users would be given by $|{\bm m}\rangle_G$. Announcing publicly the bit string ${\bm m} \oplus {\bm \mu}$ allows the end--users to shift the basis such that they finally hold the state $|\bm \mu\rangle_G$. Note that ${{\bm m}\oplus {\bm \mu}}$ is a random string from the point of view of local agents and an eavesdropper, and does not contain information since $\bm \mu$ is secret.  We consider now the case where the final state is mixed, i.e. $\rho_{\tilde G} = \sum_{{\bm \mu}{\bm \nu}} \lambda_{{\bm \mu}{\bm \nu}} |{\bm \mu}{\bm \nu}\rangle_{\tilde G}\langle {\bm \mu}{\bm \nu}|$, where ${\bm \mu}$ corresponds to vertices $1$ to $N$ (i.e. the particles held by the local agents), while ${\bm \nu}$ refers to vertices $N+1$ to $2N$ (i.e. the particles held by $C$). In the case where the outcome of all measurements is (+1), we have that the state after the measurements in $C$ is given by
\be
\rho_{\bm 0}= \sum_{\bm \mu} (\sum_{\bm \nu} \lambda_{{\bm \mu}{\bm \nu}}) |{\bm \mu}\rangle_G\langle {\bm \mu}|,
\ee
while for other measurement outcomes ${\bm m}$ the basis is shifted by ${\bm m}$, i.e. we obtain $\rho_{\bm m}$. Note that the fidelity $_G\langle {\bm m}|\rho_{\bm m}|{\bm m}\rangle_G$ with respect to the graph state corresponding to the graph $G$ is larger than the fidelity of the initial (enlarged) state with respect to the graph state $|{\bm 0}{\bm 0}\rangle_{\tilde G}$ which is given by $\lambda_{{\bm 0}{\bm 0}}$.

\subsubsection{Secturity of purification of enlarged states}

We now discuss the security of this protocol. We have that the reduced density operator of particles $1$ to $N$ held by the local agents (or an eavesdropper) is given by $\sum_{\bm m} |{\bm m} \rangle_G\langle {\bm m}|$ which is the identity for {\em any} graph $G$. This implies that even if several copies of the state are available to the local agents, they are not able to distinguish between different graphs. In fact, the state $|{\bm 0}\rangle_{\tilde G}$ (Eq. \ref{psi0}) is maximally entangled between systems $C$ and the local agents. This implies that by an appropriate unitary operation in $C$, one can change the sub--graph $G$ in $A$ to any other sub--graph $G'$. This can be seen from the fact that for a maximally entangled state $|\Phi\rangle_{CA}$, $U_C^{T}\otimes \eins_A |\Phi\rangle =\eins_A \otimes U_A |\Phi\rangle$. In other words, $C$ has in principle the freedom to change the delivered state until the last moment i.e. after the last copy is delivered to the end--users, which clearly makes it impossible for the local agents or any eavesdropper to determine the state. For these states it is also irrelevant whether the local agents learn if the final state is the result of several successful purification rounds. While in the purification of the initial graph state $|{\bm 0}\rangle_G$, this information (together with the knowledge of basis) would have allowed the local agents to exclude certain graphs (as they are not compatible with the values for correlation operators $K_j^{G}$ calculated from the measurement outcomes), here the value of each correlation operator $K_j^{\tilde G}$ depends on the unknown, random measurement outcome of a qubit in $C$ and hence is unknown (and completely random) from the point of view of the local agents or eavesdropper. Hence, one can in this case deliver only a single copy of the state to the end--users (thereby revealing that this copy is the result of several successful purification rounds) as well as only combine copies resulting from a previous successful purification step without compromising security  of the protocol.

The only possibility left to the eavesdropper to gain information about the graph is the indirect attack outlined in description of protocol ${\bf (ii)}$, Sec. \ref{Direct}. That is, Eve may adjust the noise in channels and local operations in such a way that only certain states can be purified (group 1), while others can not (group 2). From the fact whether the purification protocol was successful or not, one bit of information about the graph could be obtained in this way (namely whether the graph belongs to group 1 or 2). However, this approach necessarily forces Eve to modify channel noise and/or noise in local control operations, which can be detected by the central party $C$. To this aim, $C$ uses an enhanced validation procedure which in this case serves not only to guarantee the fidelity of the states to be purified, but is aimed to detect any possible attempts of the eavesdropper to modify the noise processes. This can be done e.g. by sending parts of maximally entangled states as probe states at certain instances instead of multipartite states to be purified (the position of these probe states remains unknown to eavesdropper). Measurements of observables $\sigma_x$ and $\sigma_z$ ---which occur naturally in the purification protocol--- can be used to perform channel tomography (at early instances of the protocol) and process tomography (at later stages of the protocol) and hence to detect any attempts of the eavesdropper to increase noise. Note that it is sufficient to send parts of maximally entangled state through channels to perform complete channel tomography \cite{Ci00}. Measurements of the two observables are sufficient in this case, as additional basis changes (and hence effective measurements of other observables) can be enforced by $C$ at the local agents during the initial channel depolarization. Local agents can in this case not distinguish whether the requested unitary operations corresponds to channel depolarization procedure or to a real basis change. The procedure is aborted if anything is not as expected (i.e. additional noise in channels or local operations occurred), even if the purification of $|{\bm 0}\rangle_{\tilde G}$ was successful.
 
Alternatively (or additionally), $C$ may prepare and purify not only the enlarged graph state $|{\bm 0}\rangle_{\tilde G}$, but also several copies of other graph states $|{\bm 0}\rangle_{\tilde G'}$ corresponding to different graphs $\tilde G'$ (with different threshold values). The positions of these states are randomly chosen. In the validation step $C$ checks if the purification of these probe states was successful. This allows $C$ to draw conclusions about attempts of Eve to introduce additional noise and eventually to abort the procedure. We remark that one typically has a hierarchy of states with respect to their fragility to noise. That is, if the purification of a certain state was successful, one can conclude that also several other states (corresponding to different graphs) are purificable. In particular, $C$ can be sure that Eve cannot distinguish between the graph $G$ and any graph $\tilde G'$ for which purification succeeded. 
We remark that the same procedure can also be used in the context of protocol {\bf (ii)}.

\section{Summary and conclusions}\label{Summary}
The purified states, created by one of the procedures {\bf (i)}, {\bf (ii)}, {\bf (iii)} serve as a resource for a variety of (security) tasks. Graph states are for instance an algorithmic specific resource, i.e. depending on which state is produced a different quantum algorithm can be applied by a simple sequence of local measurements \cite{Ra01b}. Other possible applications include secure evaluation of a (secret) function, secret sharing among some of the parties and multipartite voting schemes. The tasks that can be performed strongly depend on the structure of the graph state being produced, and thus remain unknown to any outsider party. 

We have introduced the distribution of an unknown multipartite entangled state with high fidelity as a basic quantum primitive, which can be accomplished in a real--world scenario where quantum channels as well as local control operations are noisy. We have presented three alternative ways to achieve this aim, based on channel purification and teleportation or direct multipartite entanglement purification, respectively. While the first approach is conceptually simpler, the second and third offer advantages with respect to reachable fidelities and tolerable errors.

H.-J. Briegel would like to thank J. Oppenheim for discussions.
This work was supported by the FWF, the European Union (IST-2001-38877, IST-2001-39227, IST-2004-15714), the DFG and the \"Osterreichische Akademie der Wissenschaften through project APART (W.D.).

%%%%%%%%%%%%%%%%%%%%%%%%%%%%%%%%%%%%%%%%%%%%%%%%%%%%%%%%%%%%%%%%%%%%%%%%%%%%%%


\begin{thebibliography}{99}
\bibitem{notekey}
(One of) the customers has to secretly convey the order of $G$ and ${\bm \mu}$ to the central station $C$, e.g. by using a secure classical channel. One can hence assume that an additional secret bit string can be exchanged along with the order.
\bibitem{noteensemble}
An eavesdropper might learn (a slight amount) of information about the actual value of ${\bm m_i}$ at the expense of disturbing the states, however it is impossible to learn any information about the graph $G$. As the graph $G$ specifies the (entanglement) properties of the state, the created resource remains unknown to any eavesdropper or distrustful local agents.
\bibitem{notedepol}
The diagonal coefficients of the resulting state are, in general, different for the two depolarization procedures. For the widely used class of Pauli channels, the diagnonal coefficients are the same and in fact no depolarization is required.
\bibitem{notechannel}
One arrives at exactly the same conclusions in the situation where the channels are not independent, i.e. ${\cal E}\neq {\cal E}_1{\cal E}_2\ldots {\cal E}_N$,  which could arise for example during an eavesdropping attack.
\bibitem{noteCNOT}
The CNOT operation is defined by $|i\rangle_A|j\rangle_B \rightarrow |i\rangle_A|i\oplus
j\rangle_B$.
\end{thebibliography}
\end{document}